\pdfoutput=1
\documentclass[prb,twocolumn,showpacs,amsmath,amssymb,superscriptaddress,floatfix]{revtex4}

\usepackage{graphicx}
\usepackage{dcolumn}
\usepackage{bm}
\usepackage{epsfig}

\begin{document}

\title{Microwave-driven ferromagnet--topological-insulator heterostructures: The prospect for giant spin battery effect and quantized charge pump devices}

\author{Farzad Mahfouzi}
\affiliation{Department of Physics and Astronomy, University
of Delaware, Newark, DE 19716-2570, USA}
\author{Branislav K. Nikoli\' c}
\affiliation{Department of Physics and Astronomy, University
of Delaware, Newark, DE 19716-2570, USA}
\affiliation{Department of Physics, National Taiwan University, Taipei 10617, Taiwan}
\author{Son-Hsien Chen}
\email{d92222006@ntu.edu.tw}
\affiliation{Department of Physics and Astronomy, University
of Delaware, Newark, DE 19716-2570, USA}
\affiliation{Department of Physics, National Taiwan University, Taipei 10617, Taiwan}
\author{Ching-Ray Chang}
\email{crchang@phys.ntu.edu.tw}
\affiliation{Department of Physics, National Taiwan University, Taipei 10617, Taiwan}

\begin{abstract}
We study heterostructures where a two-dimensional topological insulator (TI) is attached to two normal metal (NM) electrodes
while an island of a ferromagnetic insulator (FI) with precessing magnetization covers a portion of its lateral edges to induce time-dependent exchange field underneath via the magnetic  proximity effect. When the FI island covers both lateral edges, such device pumps {\em pure} spin current in the absence of any bias voltage, thereby acting as an efficient spin battery with giant output current even at very small microwave power input driving the precession. When only one lateral edge is covered by the FI island, both charge and spin current are pumped into the NM electrodes. We delineate conditions for the corresponding conductances (current-to-microwave-frequency ratio) to be quantized in a wide interval of precession cone angles, which is robust with respect to weak disorder and can be further extended by changes in device geometry.
\end{abstract}


\maketitle

\section{Introduction}

The recent experimental confirmation of two- (2D) and three-dimensional (3D) topological insulators~\cite{Hasan2010} (TIs), such as HgTe/(Hg,Cd)Te quantum wells~\cite{Konig2007,Roth2009} of certain width and compounds involving bismuth,~\cite{Hasan2010} respectively, has attracted considerable attention from both basic and applied research communities. The TIs introduce an exotic quantum state of matter brought by spin-orbit (SO) coupling effects in solids which is characterized by a topological invariant that is insensitive to microscopic details and robust with respect to weak disorder.~\cite{Hasan2010} Thus, although TIs have energy gap in the bulk,  their topological order  leads to quantized physical observables in the form of the number of gapless edge (in 2D) or surface (in 3D) states modulo two---TIs have an odd number of edge (surface) states in contrast to trivial band insulators with even (i.e., typically zero) number of such states.

As regards applications, the  channeling of spin transport~\cite{Akhmerov2009a} through one-dimensional (1D) counter-propagating spin-filtered (i.e., ``helical'') edge states of 2D TIs, where the time-reversal invariance  forces electrons of opposite spin  to flow in opposite directions, opens new avenues to realize semiconductor spintronic devices based on manipulation of coherent spin states.~\cite{Awschalom2007} For example, fabrication of spin-field-effect transistor~\cite{Datta1990} (spin-FET), where spin precession in the presence of SO coupling is used to switch between on and off current state, requires to prevent entanglement of spin and orbital electronic degrees of freedom in wires with many conducting channels or different amounts of spin precession along different trajectories,~\cite{Nikoli'c2005} both of which make it impossible to achieve the perfect off state of spin-FET.

\begin{figure}
\includegraphics[scale=0.3,angle=0]{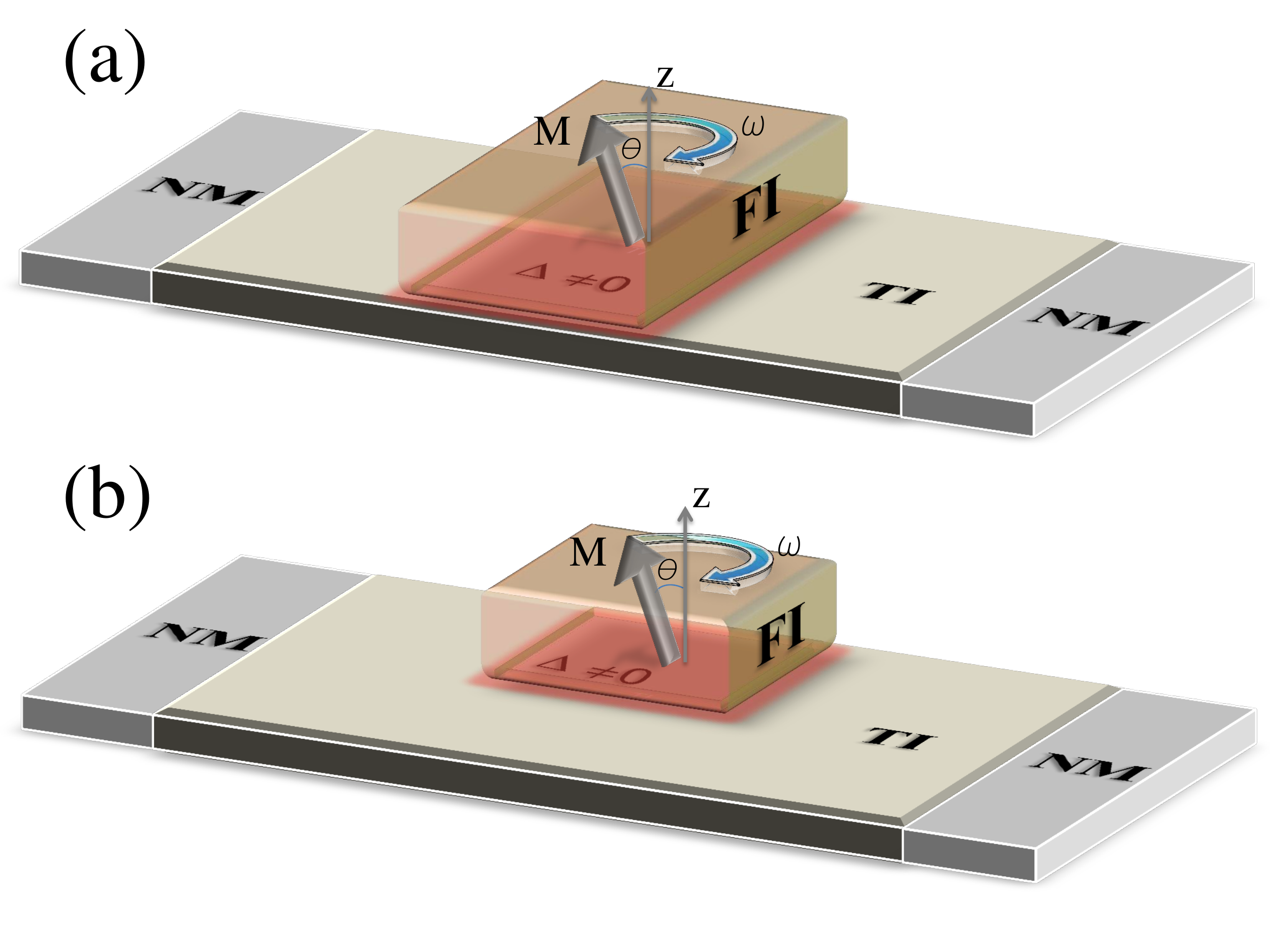}
\caption{(Color online) The proposed heterostructures consist of a 2D topological insulator (TI) attached to two normal metal (NM) electrodes where the ferromagnetic insulator (FI) with precessing magnetization (with cone angle $\theta$) under the FMR conditions induces via the proximity effect a time-dependent exchange field $\Delta \neq 0$ in the TI region underneath. In the absence of any applied bias voltage, these devices pump pure spin current into the NM electrodes in setup (a) or both charge and spin current in setup (b).}
\label{fig:setup}
\end{figure}

Some of the key questions posed by these rapid developments are: How can the TI phase be detected by conventional measurements of {\em quantized} charge~\cite{Qi2008} transport quantities? How can  spintronic {\em heterostructures}~\cite{Hasan2010} exploit TI edge or surface states in the presence of interfaces with other materials~\cite{Yokoyama2009} or internal and external magnetic fields~\cite{Yokoyama2010} used to manipulate spins while breaking the time-reversal invariance? 

For example, the 2D TI is operationally defined as a system which exhibits the quantum spin Hall effect (QSHE) with quantized spin conductance (ratio of transverse pure spin current to longitudinally applied bias voltage). However, this quantity is difficult to observe, and reported measurements~\cite{Konig2007,Roth2009} of electrical quantities probing the edge state transport in HgTe-based multiterminal devices have exhibited poor precision of quantization when contrasted with the integer quantum Hall effect---a close cousin of QSHE used in metrology.

Here we propose two ferromagnet-TI (\mbox{FM-TI}) heterostructures, illustrated in Fig.~\ref{fig:setup}, where an island of a ferromagnetic insulator (FI) is deposited over the surface of 2D TI modeled either as graphene nanoribbon (GNR)~\cite{Kane2005} with intrinsic SO coupling~\cite{Gmitra2009} or HgTe-based strip.~\cite{Konig2007,Roth2009,Bernevig2006} The precessing magnetization of FI under the ferromagnetic resonance conditions~\cite{Kajiwara2010} (FMR) will induce a time-dependent exchange field in the TI region underneath via the magnetic proximity effect.~\cite{Yokoyama2010} Using the nonequilibrium Green function (NEGF) approach~\cite{Chen2009,Chen2010,Hattori2007} to pumping by precessing magnetization in the frame rotating with it, we demonstrate that setup in Fig.~\ref{fig:setup}(a) makes possible efficient conversion of microwave radiation into {\em pure} spin current (Fig.~\ref{fig:singlefm}) whose magnitude can reach a quantized  value $eI^{S_z}/\hbar \omega = 2 \times e/4\pi$ even at small increase of the precession cone angle (i.e., microwave power input~\cite{Moriyama2008}) away from zero. On the other hand, the device in Fig.~\ref{fig:setup}(b) generates charge current $I$ (in addition to spin current) which is quantized $eI/\hbar \omega = e^2/h$ for a wide range of precession cone angles (Fig.~\ref{fig:dualfm}). This offers an alternative operational definition of the 2D TI in terms of  electrical measurements or a microwave detector which is more sensitive than conventional \mbox{FM-NM} spin pumping devices.~\cite{Moriyama2008} We also analyze the effect of disorder and device size on the quantization of pumped currents.

The paper is organized as follows. In Sec.~\ref{sec:negf}, we discuss how to compute pumped currents due to precessing magnetization by mapping such time-dependent quantum transport problem to an equivalent four-terminal DC circuit  in the frame rotating with magnetization where steady-state spin and charge currents are evaluated using NEGFs in that frame. Section~\ref{sec:spin} covers pure spin current pumping in the heterostructure of Fig.~\ref{fig:setup}(a), while Sec.~\ref{sec:charge} shows how charge current is pumped in the second type of proposed heterostructure in Fig.~\ref{fig:setup}(b). We explain the origin and the corresponding requirements for these pumped currents to be quantized in Sec.~\ref{sec:origin}. We conclude in Sec.~\ref{sec:conclusions}.

\section{Rotating frame approach to spin pumping in FM-TI heterostructures}\label{sec:negf}

The simplest model for the 2D TI central region of the device in Fig.~\ref{fig:setup} is GNR with intrinsic SO coupling, as described by  the effective single $\pi$-orbital tight-binding Hamiltonian:
\begin{eqnarray}\label{eq:hlab1}
\lefteqn{\hat{H}_{\rm GNR}^{\rm lab}(t) =   \sum_{\bf i}  \hat{c}_{\bf i}^\dag \left( \varepsilon_{\bf i} - \frac{\Delta_{\bf i}}{2} \mathbf{m}_{\bf i}(t) \cdot \hat{\bm \sigma} \right)  \hat{c}_{\bf i} } \nonumber \\
&&{} - \gamma \sum_{\langle \mathbf{ij} \rangle} \hat{c}_{\bf i}^\dag \hat{c}_{\bf j}
+ \frac{2i}{\sqrt{3}} \gamma_{\rm SO} \sum_{\langle \langle \mathbf{ij} \rangle \rangle} \hat{c}_{\bf i}^\dagger \hat{\bm \sigma} \cdot ({\bf d}_{\bf kj} \times {\bf d}_{\bf ik})\hat{c}_{\bf j}.
\end{eqnarray}
 Here $\hat{c}_{\bf i} = (\hat{c}_{{\bf i}\uparrow}, \hat{c}_{{\bf i}\downarrow})^T$ is the vector of spin-dependent operators ($\uparrow$, $\downarrow $ denotes electron spin) which annihilate electron  at site $\mathbf{i}=(i_x,i_y)$ of the honeycomb lattice, and $\hat{\bm \sigma}=(\hat{\sigma}_x,\hat{\sigma}_y,\hat{\sigma}_z)$ is the vector of the Pauli matrices. The  nearest-neighbor hopping $\gamma$ is assumed to be the same on the honeycomb lattice of GNR and square lattice of semi-infinite NM leads. The third sum in Eq.~(\ref{eq:hlab1}) is non-zero only in the GNR regions where it introduces the intrinsic SO coupling compatible with the symmetries of the honeycomb lattice.~\cite{Kane2005,Gmitra2009} The SO coupling, which is responsible for the band gap~\cite{Kane2005}  $\Delta_{\rm SO}=6\sqrt{3}\gamma_{\rm SO}$, acts as spin-dependent next-nearest neighbor hopping where ${\bf i}$ and ${\bf j}$ are two next-nearest neighbor sites, ${\bf k}$ is the only common nearest neighbor of ${\bf i}$ and ${\bf j}$, and ${\bf d}_{\bf ik}$ is a vector pointing from  ${\bf k}$ to ${\bf i}$. For simplicity,~\cite{Kane2005,Chen2010} we assume unrealistically~\cite{Gmitra2009} large value for $\gamma_{\rm SO}=0.03 \gamma$. We use the on-site potential $\varepsilon_{i} \in [-W/2,W/2]$ as a uniform random variable to model the isotropic short-range spin-independent static impurities.

In both GNR and HgTe models, the coupling of itinerant electrons to collective magnetic dynamics is described through the exchange potential $\Delta_{\bf i}$. This is assumed to be non-zero only within the region of the TI which is covered by the FI island with precessing magnetization where the proximity effect~\cite{Yokoyama2010} generates the time-dependent Zeeman term adiabatically. The magnitude of the effective exchange potential is selected to be \mbox{$\Delta=0.1\gamma$} in GNR model and \mbox{$\Delta=0.004$ eV} in HgTe model for 2D TI. The components of the rotating exchange field in the plane of the 2D TI, \mbox{$\Delta_{\bf i} m_{\bf i}^x/2$} and \mbox{$\Delta_{\bf i} m^y_{\bf i}/2$}, generate energy gap by removing the edge states from the $\Delta_{\rm SO}$ gap of the TI region below the FI island (in both models we assume $\Delta < \Delta_{\rm SO}$).

The effective tight-binding Hamiltonian~\cite{Bernevig2006} for the HgTe/CdTe quantum wells (applicable for small momenta around the $\Gamma$ point) is defined on the square lattice with four orbitals per site:
\begin{eqnarray}\label{eq:hlab2}
\hat{H}_{\rm HgTe}^{\rm lab}(t) &=&\sum_{\mathbf{i}} \hat{c}_{\mathbf{i}}^{\dagger }\left[ \left(
\begin{array}{cccc}
\varepsilon_{\mathbf{i}}^s   & 0 & 0 & 0 \\
0 & \varepsilon_{\mathbf{i}}^p  & 0 & 0 \\
0 & 0 & \varepsilon_{\mathbf{i}}^{s'}  & 0 \\
0 & 0 & 0 & \varepsilon_{\mathbf{i}}^{p'}   \\
\end{array}
\right) - \frac{\Delta_{\bf i}}{2} \mathbf{m}_{\bf i}(t) \cdot \hat{\bm \sigma} \right] \hat{c}_{\mathbf{i}}  \nonumber \\
&+&\sum_{\mathbf{i}} \hat{c}_{\mathbf{i}}^{\dagger }\left(
\begin{array}{cccc}
V_{ss} & V_{sp} & 0 & 0 \\
-V_{sp}^{\ast } & V_{pp} & 0 & 0 \\
0 & 0 & V_{ss} & V_{sp}^{\ast } \\
0 & 0 & -V_{sp}^{{}} & V_{pp} \\
\end{array}
\right) \hat{c}_{\mathbf{i} + \mathbf{e}_x} + \mathrm{H.c.}  \nonumber \\
&+&\sum_{\mathbf{i}} \hat{c}_{\mathbf{i}}^{\dagger }\left(
\begin{array}{cccc}
V_{ss} & iV_{sp} & 0 & 0 \\
iV_{sp}^{\ast } & V_{pp} & 0 & 0 \\
0 & 0 & V_{ss} & -iV_{sp}^{\ast } \\
0 & 0 & -iV_{sp} & V_{pp} \\
\end{array}%
\right) \hat{c}_{\mathbf{i}+ \mathbf{e}_y} + \mathrm{H.c.}  \nonumber \\
\end{eqnarray}%
Here vector \mbox{$\hat{c}_{\mathbf{i}}=(c_{\mathbf{i}}^s,c_{\mathbf{i}}^p,c_{\mathbf{i}}^{s'},c_{\mathbf{i}
}^{p'})^T$} contains four operators which annihilate an electron on site $\mathbf{i}$ in quantum states $|s,\uparrow \rangle
$,$ |p_{x}+ip_{y},\uparrow \rangle $ , $|s,\downarrow \rangle $, $
|-(p_{x}-ip_{y}),\downarrow \rangle$, respectively.  The Fermi energy
is uniform throughout the device in Fig.~\ref{fig:setup}, while
the on-site matrix elements,  $\varepsilon_{\mathbf{i}}^s=\varepsilon_{\mathbf{i}}^{s'}=E_s$
and $\varepsilon_{\mathbf{i}}^p=\varepsilon_{\mathbf{i}}^{p'}=E_p$, are tuned by the
gate potential to ensure that TI regions are insulating and the NM electrodes described by
the same Hamiltonian~(\ref{eq:hlab2}) are in the metallic regime. The unit vectors  ${\bf e}_x$ and ${\bf  e}_y$  are
along the $x$ and $y$ directions, respectively. The parameters
$E_{s},E_{p},V_{ss},V_{pp},V_{sp}$ characterizing the  clean HgTe/CdTe quantum wells are
defined as $V_{sp}=-iA/2a$, $
V_{ss}=(B+D)/a^{2}$, $V_{pp}=(D-B)/a^{2}$, $E_{s}=C+M-4(B+D)/a^{2}$,
and $ E_{p}=C-M-4(D-B)/a^{2}$ ($a$ is the lattice constant) where $A,B,C,D$ and $M$ are controlled experimentally.~\cite{Roth2009}

\begin{figure}
\includegraphics[scale=0.25,angle=0]{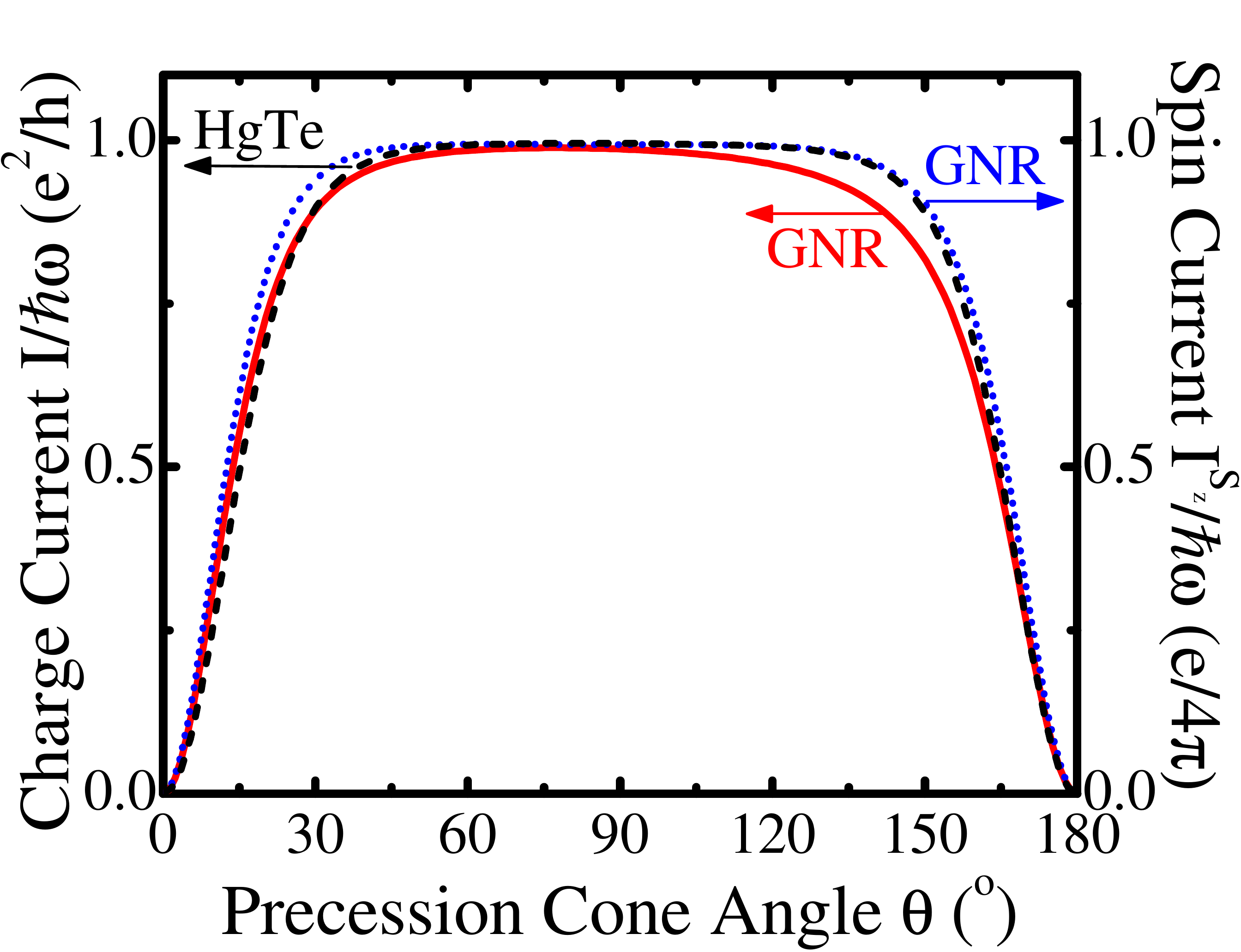}
\caption{(Color online) The total pumped charge current versus the precession cone angle in \mbox{FM-TI} heterostructures from Fig.~\ref{fig:setup}(b). The TI region is modeled as GNR with zigzag edges and intrinsic SO coupling $\gamma_{\rm SO}=0.03 \gamma$ or HgTe-based strip. In addition to charge current, these heterostructures pump spin current plotted explicitly for the GNR-based TI, while for HgTe-based device the two curves  are virtually identical (due to larger device size).}
\label{fig:dualfm}
\end{figure}

The width of GNR regions with zigzag edges is measured in terms of the number of zigzag chains $N_y$ comprising it, while its length is measured using the number of carbon atoms $d_{\rm TI}$ in the longitudinal direction.~\cite{Chen2010} The GNR-based devices studied in Figs.~\ref{fig:singlefm} and ~\ref{fig:disorder} are of the size $N_y=20$, $d_{\rm TI}=80$ where FI island of  length $d_{\rm FI}=40$ covers middle part of the TI, while in Figs.~\ref{fig:dualfm}--\ref{fig:totallocalspin} the device is smaller, $N_y=20$, $d_{\rm TI}=45$ and $d_{\rm FI}=15$, to allow for transparent images of local current profiles. The Fermi energy $E_F=10^{-6}\gamma$ is within the TI gap.

The size of HgTe-based heterostructures is measured using the number of transverse lattice sites $N_y$ and the number of sites $d_{\rm TI}$ in the longitudinal direction. The devices studied bellow have $N_y=50$, $d_{\rm TI}=200$ with FI island of length \mbox{$d_{\rm FI}=100$} covering middle part of the TI region (Fig.~\ref{fig:singlefm} also shows result for a larger device, $N_y=100$, $d_{\rm TI}=400$ and \mbox{$d_{\rm FI}=200$}).

Hamiltonians~(\ref{eq:hlab1}) and (\ref{eq:hlab2}) are time-dependent since the spatially uniform unit vector ${\bf m}(t)$ along the local magnetization direction is precessing  steadily around the $z$-axis with a constant precession cone angle $\theta$ and frequency $f=\omega/2\pi$. This complicated time-dependent transport problem can be transformed into a simpler time-independent one via the unitary transformation of Hamiltonians (\ref{eq:hlab1}) or (\ref{eq:hlab2}) using $\hat{U}=e^{i \omega \hat{\sigma}_z t/2}$  [for ${\bf m}(t)$
precessing counterclockwise]:
\begin{equation}\label{eq:hrot}
\hat{H}_{\rm rot}  =  \hat{U} \hat{H}^{\rm lab}(t) \hat{U}^{\dagger} - i \hbar \hat{U} \frac{\partial}{\partial t} \hat{U}^\dagger = \hat{H}^{\rm lab}(0)  - \frac{\hbar \omega}{2} \hat{\sigma}_z.
\end{equation}
The Zeeman term $\hbar \omega\hat{\sigma}_z/2$, which emerges uniformly in the sample and NM electrodes, will spin-split the bands of the NM electrodes, thereby providing a rotating frame picture of pumping based on the four-terminal DC device.~\cite{Chen2009,Chen2010,Hattori2007}  In the equivalent DC device, pumping by precessing magnetization can be understood~\cite{Chen2009} as a flow of spin-resolved charge currents between four spin-selective (i.e., effectively half-metallic FM) electrodes $_L^\downarrow$, $_L^\uparrow$, $_R^\downarrow$ $_R^\uparrow$ ($L$-left, $R$-right) biased by the electrochemical potential difference $\mu^\downarrow_p - \mu^\uparrow_p = \hbar \omega$.

The basic transport quantity for the DC circuit in the rotating frame is the spin-resolved bond charge current carrying spin-$\sigma$ electrons from site $\mathbf{i}$ to site $\mathbf{j}$
\begin{equation}\label{eq:jlab}
J_{\bf ij}^\sigma = \frac{e}{h} \int_{-\infty}^{\infty} dE\, [\gamma_{\bf ij} \bar{G}^{<,\sigma \sigma}_{\bf j i}(E) - \gamma_{ji}\bar{G}^{<,\sigma\sigma}_{\bf ij}(E)].
\end{equation}
This is computed in terms of the lesser Green function in the rotating frame~\cite{Chen2009,Chen2010,Hattori2007} $\bar{G}^<(E)$. Unlike  $G^<(t,t^\prime)$ in the lab frame, $\bar{G}^<$ depends on only one time variable $\tau=t-t^\prime$ (or energy $E$ after the time difference $\tau$ is Fourier transformed). This finally yields local spin
\begin{equation}\label{eq:localspin}
J^S_{\bf ij}=\frac{\hbar}{2e}\left(J^{\uparrow}_{\bf ij}-J^{\downarrow}_{\bf ij} \right),
\end{equation}
and local charge
\begin{equation}\label{eq:localcharge}
J_{ij}=J^{\uparrow}_{\bf ij}+J^{\downarrow}_{\bf ij},
\end{equation}
currents flowing between nearest neighbor or next-nearest neighbor sites $\mathbf{i}$ and $\mathbf{j}$ connected by hopping $\gamma_{\bf ij}$. They can be computed within the device or within the NM electrodes.

The summation of all $J^S_{\bf ij}$ or $J_{\bf ij}$ at selected transverse cross section, \mbox{$I^{S_z} = \sum_{ij} J_{ij}^{S_z}$} (assuming the $z$-axis for the spin quantization axis) and \mbox{$I = \sum_{ij} J_{ij}$}, yields total spin or charge current, respectively. The charge current $I$ has to be the same at each cross section due to charge conservation, but the spin current $I^{S_z}$ can vary in different regions of the device since spin does not have to be conserved. The magnitude of total currents pumped into, e.g., the left NM electrode (i.e., computed at any cross section within the left NM electrode) can also be expressed in terms of the transmission coefficients for the four-terminal DC device in the rotating frame~\cite{Chen2009}
\begin{eqnarray}\label{eq:totalscurrent}
I^{S_z}_L & = &  \frac{e}{h} \int dE\, \left(T_{RL}^{\uparrow \downarrow} + T_{LR}^{\uparrow \downarrow}  + 2T_{LL}^{\uparrow \downarrow}\right) \nonumber \\
\displaystyle && \times \left[ f^\downarrow(E)-f^\uparrow(E)\right], \\
\label{eq:totalccurrent}
I &  = & \frac{1}{4\pi} \int dE\, \left(T_{RL}^{\uparrow \downarrow}- T_{LR}^{\uparrow \downarrow}\right) \left[f^\downarrow(E)-f^\uparrow(E)\right],
\end{eqnarray}
Here the transmission coefficients $T_{pp'}^{\sigma \sigma'}$ determine the probability for $\sigma'$ electrons injected through lead $p'$ to emerge in electrode $p$ as spin-$\sigma$ electrons, and can be expressed in terms of the spin-resolved NEGFs.~\cite{Chen2009} The distribution function of electrons in the four electrodes of the DC device is given by
\mbox{$f^{\sigma}(E)=\{\exp[(E-E_F+\sigma\hbar\omega/2)/kT]+1\}^{-1}$} where $\sigma=+$ for spin-$\uparrow$ and $\sigma=-$ for spin-$\downarrow$. Since the device is not biased in the laboratory frame (where all NM electrodes are at the same electrochemical potential $\mu_p=E_F$), this shifted Fermi function is uniquely specified by the polarization $\uparrow$ or $\downarrow$ of the  electrode.

\section{Quantized pure spin current pumping in FM-TI heterostructures} \label{sec:spin}

\begin{figure}
\includegraphics[scale=0.25,angle=0]{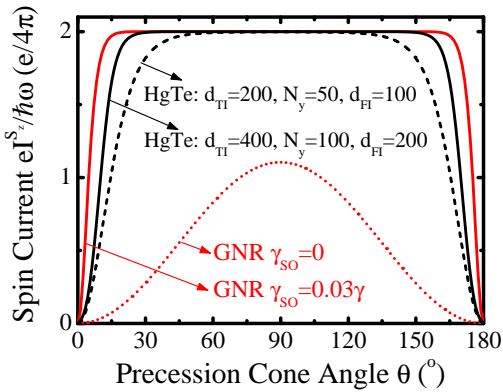}
\caption{(Color online) The total pure spin current pumped into the NM electrodes as a function of the precession cone angle in \mbox{FM-TI} heterostructures from Fig.~\ref{fig:setup}(a). The TI region is modeled as GNR with zigzag edges and non-zero intrinsic SO coupling $\gamma_{\rm SO} \neq 0$ or HgTe-based strip. For comparison, we also plot pumped spin current when TI is replaced by a zigzag GNR with zero intrinsic SO coupling $\gamma_{\rm SO}=0$. In the case of HgTe-based heterostructure, we show that increasing the size of the proximity induced magnetic region within TI widens the interval of cone angles within which pumped current is quantized.}
\label{fig:singlefm}
\end{figure}

The precessing magnetization of FM island in the device setup of Fig.~\ref{fig:setup}(a) pumps pure (i.e., with no accompanying net charge flux) spin current symmetrically  into the left and right NM electrodes in the absence of any bias voltage [if the device is asymmetric, charge current is also pumped but only as the second order $\propto (\hbar\omega)^2$ effect~\cite{Chen2009}]. In the case of conventional NM in contact with precessing FM, different approaches predict~\cite{Brataas2002,Chen2009} that pumped spin current by the FM$|$NM interface behaves as $I^{S_z} \propto \sin^2 \theta$. To understand the effect of the TI surrounding the precessing island, we first reproduce this feature in Fig.~\ref{fig:singlefm} for GNR with no SO coupling ($\gamma_{\rm SO}=0$). When the intrinsic SO coupling~\cite{Gmitra2009} is ``turned on'' ($\gamma_{\rm SO} \neq 0$), the pumped pure spin current in Fig.~\ref{fig:singlefm} is substantially enhanced (by up to two orders of magnitude at small precession cone angles). In fact, pumping into helical edge states {\em profoundly modifies} $I^{S_z}$ vs $\theta$ characteristics which becomes constant quantized quantity  $eI^{S_z}/\hbar \omega = 2 \times e/4\pi$ for large enough $\theta$.

Figure~\ref{fig:singlefm} also confirms the same behavior for HgTe model of 2D TI. Moreover, it shows that interval of cone angles within which pumped current is quantized can be manipulated by using longer FI region. Exploiting this feature would enable giant spin battery effect where large pure spin current is induced by even very small microwave power input which experimentally~\cite{Moriyama2008} controls the precession cone angle.

Note that since $\hbar\omega \ll E_F$, we can use \mbox{$f^\downarrow(E)-f^\uparrow(E) \approx \hbar \omega \delta(E-E_F)$} at low temperatures for the difference of the Fermi functions in Eqs.~(\ref{eq:totalscurrent}) and ~(\ref{eq:totalccurrent}). This ``adiabatic approximation''~\cite{Hattori2007} is analogous to linear response calculations for biased devices, allowing us to define the pumping spin conductance $G_{\rm SP}=eI^{S_z}/\hbar\omega$. Its quantization in Fig.~\ref{fig:singlefm} is an alternative characterization of the 2D TI phase when compared to QSHE in four-terminal bridges~\cite{Kane2005,Chen2010} where longitudinal charge current driven by the bias voltage $V$ generates transverse spin current $I_T^{S_z}$ and corresponding quantized spin Hall conductance $G_{\rm SH}=I_T^{S_z}/V = 2 \times e/4\pi$. Thus, the spin battery in Fig.~\ref{fig:setup}(a) would produce much larger pure spin currents than currently achieved  through, e.g., conventional SHE in low-dimensional semiconductors while offering tunability that has been difficult to demonstrate for SHE-based devices.~\cite{Awschalom2007}

\begin{figure}
\includegraphics[scale=0.4,angle=0]{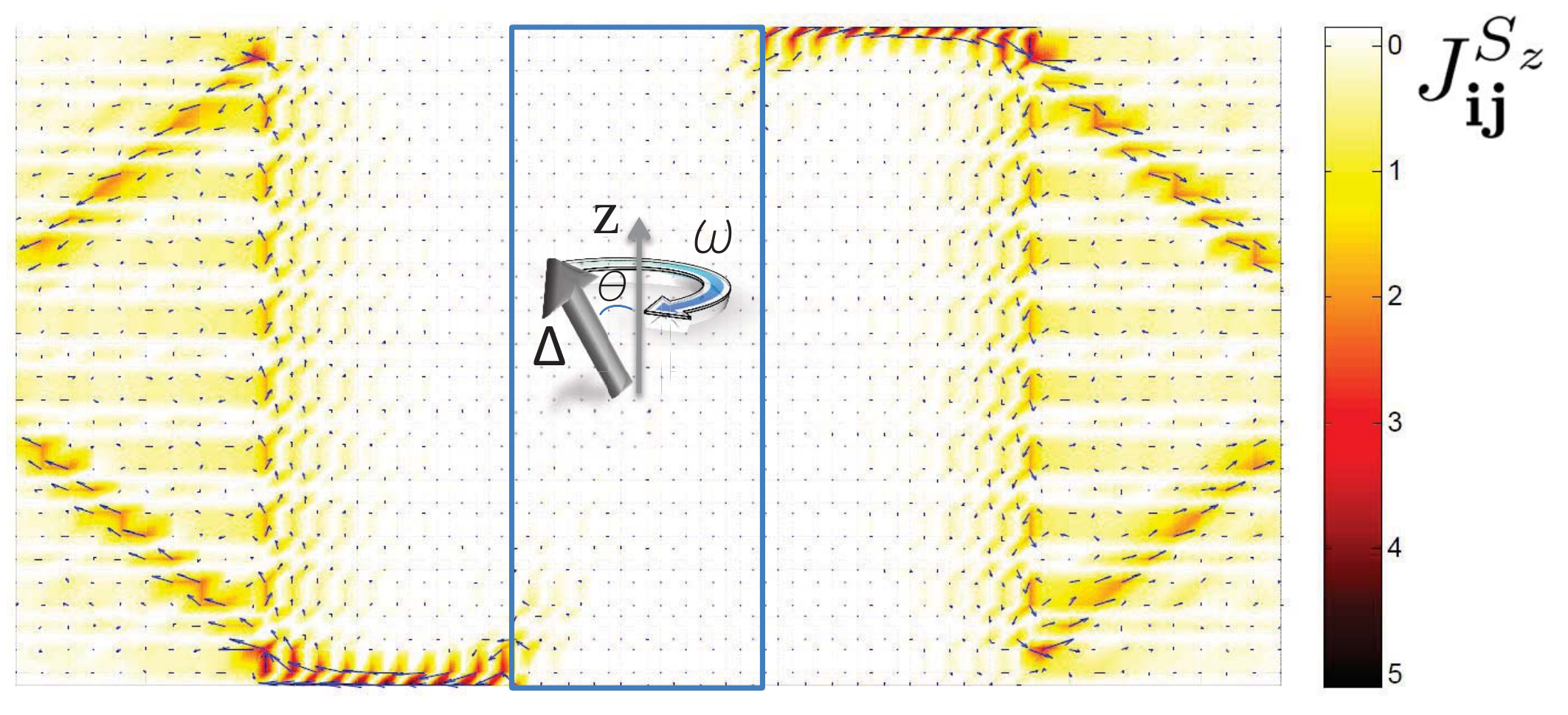}
\caption{(Color online) Spatial profile of local pumped pure spin current corresponding to total current in
Fig.~\ref{fig:singlefm} at $\theta = 90^\circ$ for GNR model of TI with $\gamma_{\rm SO} = 0.03 \gamma$. The corresponding total pumped currents are plotted in Figs.~\ref{fig:singlefm} and ~\ref{fig:totallocalspin}.}
\label{fig:localspin}
\end{figure}

We recall that the original proposal~\cite{Brataas2002} for spin battery operated by FMR was based on \mbox{FM-NM} heterostructures. However, experiments~\cite{Gerrits2006} performed on Ni$_{80}$Fe$_{20}|$Cu bilayers have found that spin pumping by FM$|$NM interfaces is not an efficient scheme to drive spin accumulation in nonmagnetic materials (e.g., estimated~\cite{Gerrits2006} spin polarization is only $2 \times 10^{-6}$ in 10-nm-thick Cu layer) because of the backflow of accumulated spins into the FM and the diffusion of polarized spins inside the NM. No such spin accumulation or spin dephasing exists in the device in Fig.~\ref{fig:setup}(a) where bulk transport within the TI regions is completely suppressed (see Fig.~\ref{fig:localspin}) while 1D spin transport is guided by helical edge states.

\section{Quantized charge current pumping in FM-TI heterostructures} \label{sec:charge}

While the most direct confirmation of the 2D TI phase would be achieved by measuring quantized $G_{\rm SH}$, this is very difficult to perform experimentally. Thus, several recent studies~\cite{Qi2008,Chen2010} have proposed experiments that would detect edge state transport in  2D TIs via simpler measurement of conventional electrical quantities in response to external probing fields.

In particular, Ref.~\onlinecite{Qi2008} has conjectured that a setup with two disconnected FM islands covering two lateral edges of 2D TI,  where the magnetization of one of them is precessing while the other one is static, could pump quantized charge current counting the number of helical edge states. This proposal, based on intuitive arguments~\cite{Qi2008} rather than full quantum transport analysis of adiabatic pumping, concludes that charge pumping conductance $G_{\rm CP}=eI/ \hbar \omega = e^2/h$ would be `universally' quantized  for arbitrary device parameters or precession cone angle.

In order to induce quantized charge current response from the 2D TI phase, we propose an alternative heterostructure in Fig.~\ref{fig:setup}(b) where  FI island with precessing magnetization is covering portion of a {\em single} lateral edge of the TI. Figure~\ref{fig:dualfm} demonstrates that this device pumps both charge and spin currents into the NM electrodes. The pumping conductances $G_{\rm CP}$ plotted in Fig.~\ref{fig:dualfm} are quantized in a wide interval of precession cone angles, which can also be expanded by using longer FI island similarly to HgTe curves in Fig.~\ref{fig:singlefm}.

\section{Origin and requirements for quantized pumping in FM-TI heterostructures}\label{sec:origin}

\begin{figure}
\includegraphics[scale=0.4,angle=0]{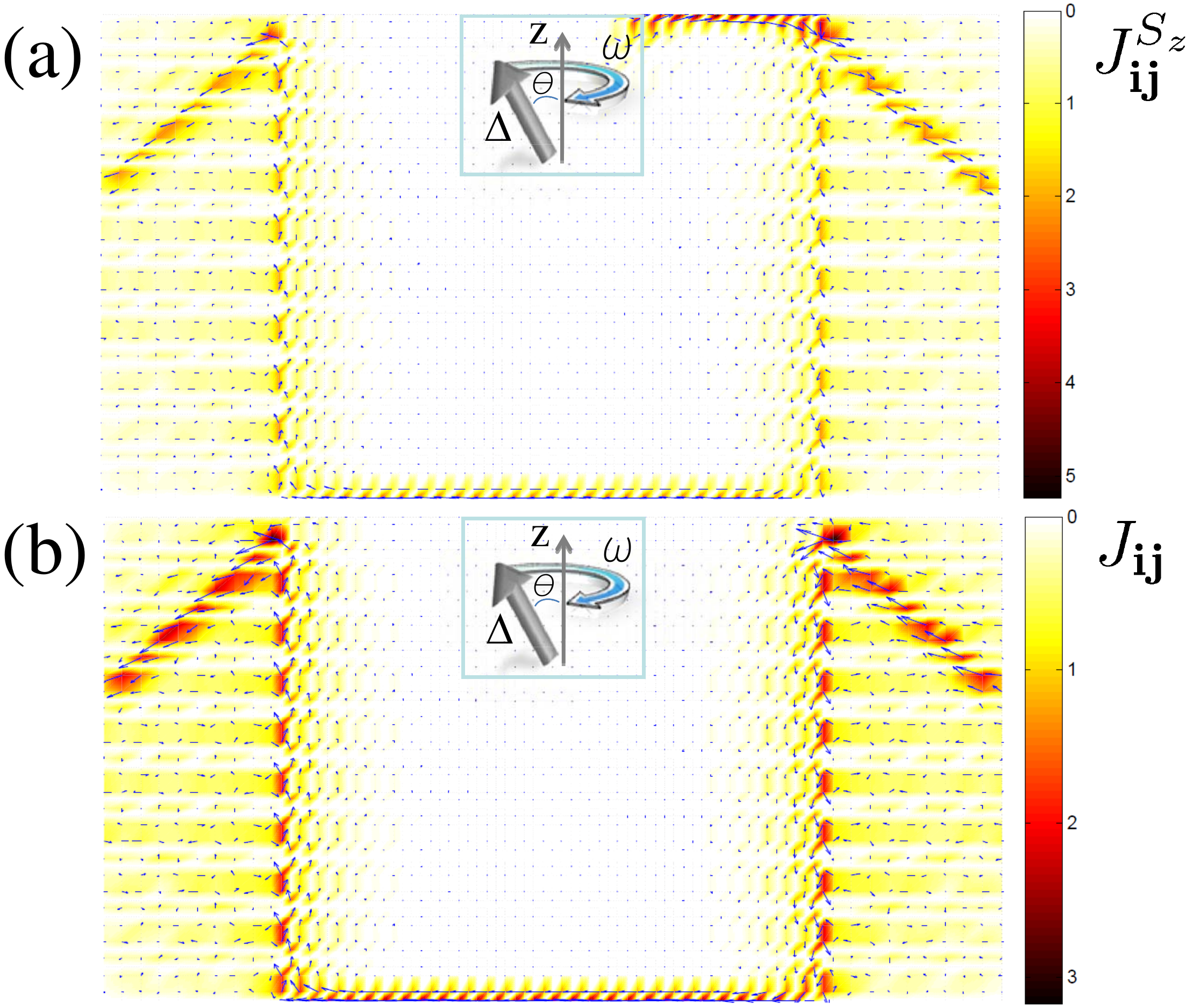}
\caption{(Color online) Spatial profile of (a) local pumped spin current and (b) local pumped charge current in the heterostructure shown in Fig.~\ref{fig:setup}(b) at $\theta = 90^\circ$. The corresponding total pumped currents are plotted in Fig.~\ref{fig:dualfm}.}
\label{fig:localcharge}
\end{figure}

To explain the origin of quantized spin and charge pumping in the proposed \mbox{FM-TI} heterostructures, we compute spatial profiles of local pure spin current in Fig.~\ref{fig:localspin}  and local charge and spin currents in Fig.~\ref{fig:localcharge} for devices in Fig.~\ref{fig:setup}(a) and Fig.~\ref{fig:setup}(b), respectively. In the four-terminal DC device picture of pumping,~\cite{Chen2009} these local {\em nonequilibrium} currents are generated by the spin flow from electrode $_p^\downarrow$ at higher $\mu_p^\downarrow$ into electrode  $_{p'}^\uparrow$ at lower $\mu_{p'}^\uparrow$. The role of the central island with static (in the rotating frame) noncollinear magnetization, for which the incoming spins are not the eigenstates of the corresponding Zeeman term, is to allow for transmission with spin precession or reflection accompanied by spin rotation (for transport between  $_p^\downarrow$ and $_p^\uparrow$ electrodes). The spin precession or rotation is necessary for spin to be able to enter electrode at a lower electrochemical potential (accepting spins opposite to the originally injected ones) while flowing through the edge state moving in proper direction compatible with their chirality.

\begin{figure}
\includegraphics[scale=0.25,angle=0]{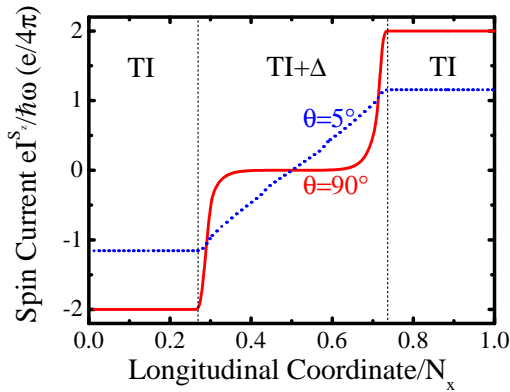}
\caption{(Color online)  Total pure spin current at each transverse cross section along the heterostructure in Fig.~\ref{fig:setup}(a) for two different precession cone angles. The total spin current for cone angle
$\theta = 90^\circ$ is obtained by summing local currents shown in Fig.~\ref{fig:localspin}.}
\label{fig:totallocalspin}
\end{figure}

The quantization of the pumped pure spin current in Fig.~\ref{fig:singlefm} is ensured by the absence of flow through the bulk of the magnetic island within TI underneath FI in Fig.~\ref{fig:localspin}(a). In this case, only perfect reflection with spin rotation at the interface between TI region with proximity induced $\Delta \neq 0$ and TI itself takes place redirecting spins from one helical edge state to the other one at the same edge. Thus, the transmission coefficient~\cite{Chen2009} $T_{LL}^{\uparrow \downarrow}=1$ in Eq.~(\ref{eq:totalscurrent}) becomes quantized since it is governed by local ballistic transport through edge states on the top right lateral edge in Fig.~\ref{fig:localspin}(a), while the other two coefficients are zero $T_{RL}^{\uparrow \downarrow} =  T_{LR}^{\uparrow \downarrow}=0$. This also explains why the range of precession cone angles within which $G_{\rm SP}$ in Fig.~\ref{fig:singlefm} or $G_{\rm CP}$  in Fig.~\ref{fig:dualfm} is quantized can be expanded by increasing the length of the magnetic island within TI (i.e., the corresponding FI island on the top) or the proximity induced exchange potential $\Delta$---both tunings diminish overlap of evanescent modes from the two TI$|$magnetic-island interfaces. This is further clarified by Fig.~\ref{fig:totallocalspin} where spin current emerges also in the bulk of the magnetic island in the non-quantized case for small $\theta=5^\circ$. As discussed in Sec.~\ref{sec:negf}, spin current is in general not conserved, which is exemplified in Fig.~\ref{fig:totallocalspin} by  different values of the total pumped spin current at different cross sections (including zero in the middle of the magnetic island at large precession cone angle $\theta = 90^\circ$; the non-zero current around interfaces is due to evanescent modes).

Analogously, quantized charge current in Fig.~\ref{fig:dualfm} is driven by the same reflection process discussed above which then generates flow of rotated
spin along the right TI$|$NM interface and the bottom lateral edge in Fig.~\ref{fig:localcharge}(b) while utilizing only one of the two helical edge states. In the lab frame picture of pumping, the emission of currents in the absence of bias voltage can be viewed as a flow of spins, driven by absorption of microwave photons, from the region around the interface between the magnetic island and TI where edge states penetrate as evanescent modes into the island. However, this framework does not offer simple explanation of why pumped currents can become quantized.

Figures~\ref{fig:localspin} and ~\ref{fig:localcharge} also provide answer to the following question: What happens to current, which is confined to a narrow region  of space along the samples edges within TI, as it exists from the TI region into the  NM electrodes?  The local charge or spin fluxes remain confined to a narrow ``flux tube''  even within the NM electrodes which is refracted at the TI$|$NM interface by an angle $45^\circ$. This feature is explained by the fact that at the TI$|$NM interface the helical edge state in the, e.g., upper right corner changes direction (to flow downward along the TI$|$NM interface) so that at this region of space at which current penetrates from TI into NM the quantum state carrying it has wavevector $k_y=k_x$. By continuity of wavefunctions, this relation is preserved within the NM electrodes leading to the observed refraction of the guiding center for electron quantum-mechanical propagation.

\begin{figure}
\includegraphics[scale=0.25,angle=0]{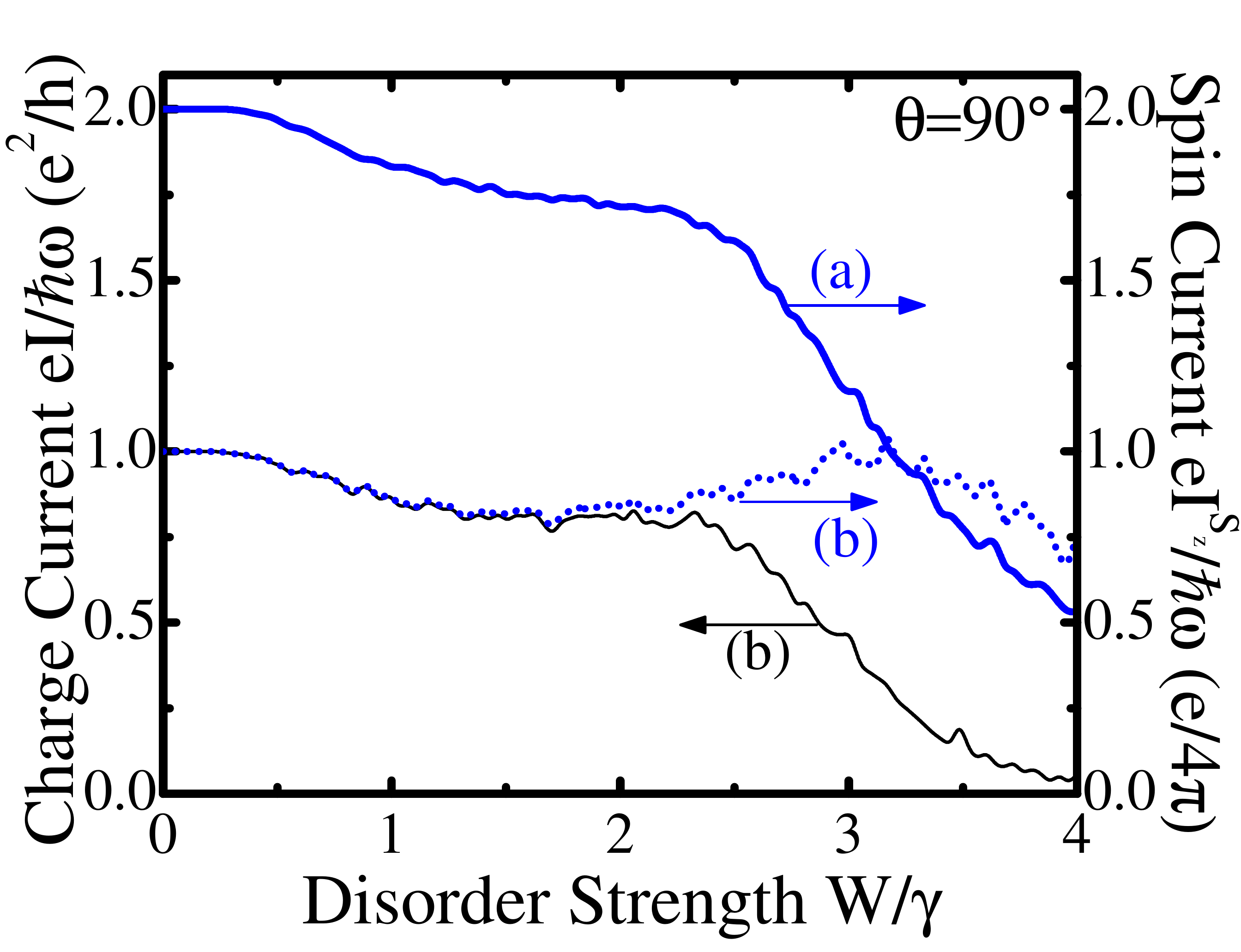}
\caption{(Color online) The effect of the static impurity potential on pumped currents at precession cone angle $\theta=90^\circ$ for GNR-based TI, where pure spin current curve labeled with (a) is generated by the spin battery device in Fig.~\ref{fig:setup}(a) while curves labeled with (b) are for the device in Fig.~\ref{fig:setup}(b).}
\label{fig:disorder}
\end{figure}

Figure~\ref{fig:disorder} shows that pumped currents remain precisely quantized in the presence of weak static (spin-independent) disorder simulating short-range impurity scattering.  Further increasing of the disorder strength  diminishes pumped charge current much faster than the spin current.

Finally, our analysis shows that the second FM island with static magnetization covering the opposite edge of the device in the proposal of Ref.~\onlinecite{Qi2008} for quantized charge pump is redundant. Moreover, in the case of FM island with precessing magnetization deposited
directly on the top of TI to generate proximity effect and pumping, quantization  would be lost~\cite{Chen2010} if electrons can penetrate
into the metallic region provided by such islands so that transport ceases to be governed purely by the helical edge states.

\section{Conclusions} \label{sec:conclusions}

In conclusion, we have proposed two types of \mbox{FM-TI} heterostructures shown in Fig.~\ref{fig:setup} which can pump quantized spin or charge current in the absence of any applied bias voltage. The device in Fig.~\ref{fig:setup}(a) emits pure spin current $I^{S_z}$ toward  both the left and the right NM electrodes. Its quantized value $eI^{S_z}/(\hbar\omega) = 2 \times e/4\pi$ can be attained even at very small microwave power input (determining the precession cone angle~\cite{Moriyama2008}) driving the magnetization precession, thereby offering a very efficient spin battery device that would surpass any battery~\cite{Brataas2002,Gerrits2006} based on pumping by conventional FM$|$NM interfaces. On the other hand, the device in Fig.~\ref{fig:setup}(b) generates quantized charge current $eI/(\hbar\omega) = e^2/h$ in response to absorbed microwaves, which can be utilized either for electrical detection of the 2D TI phase via measurement of precisely quantized quantity (that survives weak disorder) directly related to the number of helical edge states or as a sensitive detector of microwave radiation.

\begin{acknowledgments}
Financial support through NSF Grant No. ECCS 0725566 is gratefully acknowledged. C.-R. Chang was supported by the Republic of China National Science Council Grant No. \mbox{95-2112-M-002-044-MY3}.
\end{acknowledgments}



\end{document}